\documentclass{vgtc}                          % final (conference style)
\ifpdf%                                % if we use pdflatex
  \pdfoutput=1\relax                   % create PDFs from pdfLaTeX
  \usepackage{graphicx}                % allow us to embed graphics files
  \DeclareGraphicsExtensions{.pdf,.png,.jpg,.jpeg} % for pdflatex we expect .pdf, .png, or .jpg files
\else%                                 % else we use pure latex
  \usepackage{graphicx}                % allow us to embed graphics files
  \DeclareGraphicsExtensions{.eps}     % for pure latex we expect eps files
\fi%

\graphicspath{{figures/}{./}} % where to search for the images

\usepackage{microtype}                 
\PassOptionsToPackage{warn}{textcomp}  
\usepackage{textcomp}                  
\usepackage{mathptmx}                  
\usepackage{times}                     % we use Times as the main font
\usepackage{cite}                      % needed to automatically sort the references
\usepackage{tabu}                      % only used for the table example
\usepackage{booktabs}                  % only used for the table example
\usepackage{txfonts}
\usepackage{color}
\usepackage{svg}
\usepackage{paralist}
\usepackage{ccicons}     % Cite your images correctly!

\usepackage{tabularx}
\usepackage{csquotes}

\usepackage{pgfplots}
\pgfplotsset{width=\columnwidth, compat=1.8}
\usepgfplotslibrary{statistics}

\usepackage[shortlabels]{enumitem}

\usepackage{tkz-kiviat} 
\usetikzlibrary{arrows}
\usepackage{adjustbox}

\usepackage{flushend}

\newcommand{\emptypie}[1][1ex]{\tikz\draw (0,0) circle (#1);} 
\newcommand*\halfpie[1][1ex]{%
  \begin{tikzpicture}
  \draw[fill] (0,0)-- (90:#1) arc (90:270:#1) -- cycle ;
  \draw (0,0) circle (#1);
  \end{tikzpicture}}
\newcommand*\fullpie[1][1ex]{\tikz\fill (0,0) circle (#1);}

\onlineid{1016}

\vgtccategory{Technical Papers}

\usepackage{tikz}
\renewcommand\copyrighttext{%
  \tiny \textcopyright 2020 IEEE. Personal use of this material is permitted. Permission from IEEE must be obtained for all other uses, in any current or future media, including reprinting/republishing this material for advertising or promotional purposes, creating new collective works, for resale or redistribution to servers or lists, or reuse of any copyrighted component of this work in other works. Cite this article as follows: M. Beran, F. Hrdina, D. Kou\v{r}il, R. O\v{s}lej\v{s}ek and K. Z\'akop\v{c}anov\'a, "Exploratory Analysis of File System Metadata for Rapid Investigation of Security Incidents," \textit{2020 IEEE Symposium on Visualization for Cyber Security (VizSec)}, Salt Lake City, USA, 2020, pp. 11-20, doi: \url{https://doi.org/10.1109/VizSec51108.2020.00008}.
  }
\newcommand\copyrightnotice{%
\begin{tikzpicture}[remember picture,overlay]
\node[anchor=south,yshift=12pt] at (current page.south) {\fbox{\parbox{\dimexpr\textwidth-\fboxsep-\fboxrule\relax}{\copyrighttext}}};
\end{tikzpicture}%
}

\title{Exploratory Analysis of File System Metadata for Rapid Investigation of Security Incidents}

\author{
 Michal Beran\thanks{e-mail: beran@ics.muni.cz}
\and Franti\v{s}ek Hrdina\thanks{e-mail: hrdina@ics.muni.cz}
\and Daniel Kou\v{r}il\thanks{e-mail: kouril@ics.muni.cz}
\and Radek O\v{s}lej\v{s}ek\thanks{e-mail: oslejsek@fi.muni.cz}
\and Krist\'ina Z\'akop\v{c}anov\'a\thanks{e-mail: zakopcanova@mail.muni.cz}
}
\affiliation{\scriptsize Masaryk University}

\teaser{
  \centering
  \includegraphics[width=\textwidth]{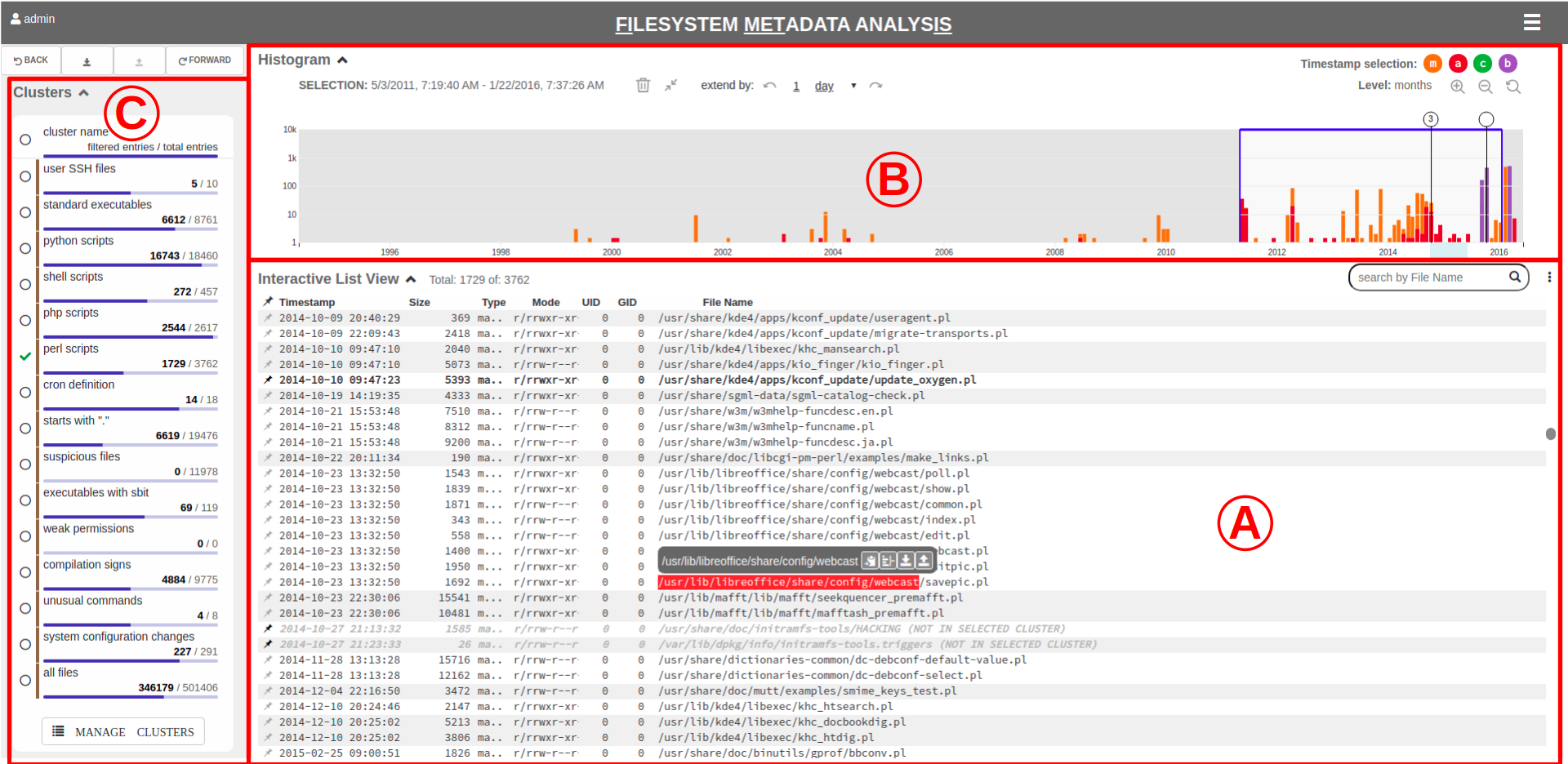}
  \caption{FIMETIS is a tool providing an interactive exploration of file system snapshots. Analysts can quickly investigate cybersecurity incidents via three complementary views: A -- \emph{list view} with file system records, B -- \emph{histogram} with a timeline, and C -- data \emph{clusters}.}
  \label{fig:teaser}
}

\abstract{
Investigating cybersecurity incidents requires in-depth knowledge from the analyst. Moreover, the whole process is demanding due to the vast data volumes that need to be analyzed. While various techniques exist nowadays to help with particular tasks of the analysis, the process as a whole still requires a lot of manual activities and expert skills. We propose an approach that allows the analysis of disk snapshots more efficiently and with lower demands on expert knowledge. Following a user-centered design methodology, we implemented an analytical tool to guide analysts during security incident investigations. The viability of the solution was validated by an evaluation conducted with members of different security teams.
}


\CCScatlist{
  \CCScatTwelve{Human-centered computing}{Visual analytics}{}{};
  \CCScatTwelve{Security and privacy}{Systems security}{File system security}{};
  \CCScatTwelve{Applied computing}{Computer forensics}{Evidence collection, storage and analysis}{};
}

\begin{document}
\maketitle
\copyrightnotice

\section{Introduction}

Cybercrime has rapidly developed over the past years~\cite{gartner}, and
cybersecurity threats are expected to present significant risks for the
future~\cite{Anderson2019}. For computer systems to be able to face the constantly changing threat
landscape, it is necessary to develop and maintain capabilities for
responding to cybersecurity attacks. A vital part of the response process consists of the investigation of the evidence, which reveals the nature of 
the incident and performed activities.

The investigation depends heavily on a proper evaluation of all collected
evidence. Methods of digital forensics~\cite{casey2009,kavrestad2018fundamentals} are
employed for systematic scrutiny of the data. It is a continuous process where hypotheses are formulated based on observations followed by steps to either confirm or deny the theory. 

\begin{figure*}[!htbp]
  \centering
  \includegraphics[width=.8\textwidth]{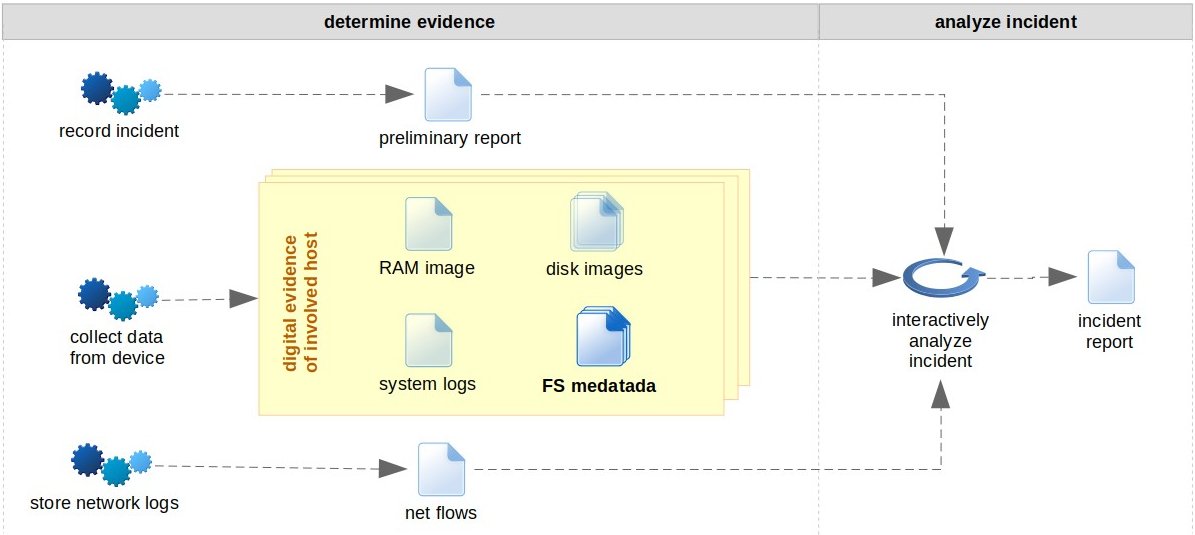}
  \caption{Incident investigation process. The \emph{FIMETIS} tool deals with file system metadata only.}
  \label{fig:workflow}
\end{figure*}

A simplified scheme of an investigation workflow is depicted in Figure~\ref{fig:workflow}. First, the suspicion of an incident is reported in the form of a preliminary report. Then, data sources for digital evidence of the incident are collected. They capture either the broader state of involved computer networks and communication history (net flows, PCAPs) or the state of involved devices (system logs, the content of disks, memory snapshots, etc.).

The iterative investigation is often time-consuming and requires a high level of expert
knowledge. The amount of data collected is often high, which only
complicates the analysis. While the forensic investigation methods 
provide a great platform to derive particular results, a user-oriented approach is missing to simplify the overall process. %As the number of incident investigations grows, a review of analysis methods is necessary to improve the efficiency of the analysis.

Permanent storage devices are a crucial part of contemporary computer systems and data retrieved from these devices provide significant input for the investigation. The state of permanent storage can be captured in multiple ways. The most straightforward and complete approach is to analyze the complete disk content. However, as current media tend to be quite large---it is not uncommon for disks to provide several terabytes of capacity---the analysis becomes time- and resource-demanding. Moreover, analyzing disk content encounters privacy issues when the data contain sensitive information~\cite{forensicsChallenges}.

One way of coping with the volume and privacy problems is to work only with file metadata, extracted from permanent storage, which include the file owner, size, name and dates of last manipulations.
However, even though such a dataset is much smaller in size compared to raw disk images, it is still necessary to process hundreds of thousands of records already in case of a standard storage. Moreover, it requires deep knowledge about the relationships among files, their purpose in the system, and importance for the attacker.

In this paper, we propose visual-analytic methods that make the investigation of file system metadata significantly more efficient and are also available to analysts with no deep domain knowledge. We describe an application called \emph{FIMETIS} (FIlesystem METadata analysIS) that was developed to verify the visual-analytic concepts. Evaluation of this tool has shown that the user interface is easy to learn and well supports analytical tasks. Even less skilled participants were able to investigate and reconstruct a real incident in limited time at surprising precision and level of details.

\section{Related Work}

Many tools and approaches dealing with individual types of data sources for digital evidence can be found. 

So far, big attention has been paid to the investigation of network communication. NetCapVis~\cite{ulmer2019} provides a post-incident visual analysis of PCAP files that capture network traffic. TVi~\cite{boschetti2011} is a tool that combines multiple visual representations of network traces to support different levels of visual-based querying and reasoning required for making sense of complex traffic data. Visualization techniques proposed by Gray et al.~\cite{gray2015} provide conceptual network navigation for situational awareness in network communication.

Analysis of system logs was researched as part of ELVIS~\cite{humphries2013} and CORGI~\cite{humphries2014}, for instance. These tools, both proposed by the same authors, provide security-oriented log visualizations that allow security experts to visually explore and link numerous types of log files through relevant representations and global filtering. A top-down approach to the log exploration is provided by the Visual Filter~\cite{stange2014} tool, which represents the whole log in a single overview and then allows the investigators to navigate and make context-preserving sub-selections.

Disks and permanent storage provide another valuable source of information for the digital investigation. Disk and file systems analysis can be performed in several layers~\cite{carrier2005}. Approaches addressing specific features are, for example, Change-link 2.0~\cite{leschke2013}, which provides several visualizations to capture changes to files and directories over time, or the work of Heitzmann et al.~\cite{heitzmann2008}, who proposed a visual representation of access control permissions in a standard hierarchical file system using treemaps. 

This paper deals with the utilization of file system metadata as they have lesser demands on volumes and do not threaten data sensitivity. 
The utility of metadata for digital forensics has been articulated previously~\cite{buchholz2004}, and various techniques for metadata-based analyses have been proposed since then.
The use of metadata to provide a fingerprint of actions performed with files has been suggested to streamline file system analysis~\cite{kalber2013}.

Metadata attributes are also known to be useful to reconstruct a timeline of previous activities~\cite{hargreaves2012} and have been demonstrated to locate suspicious files~\cite{rowe2011}. These techniques address the particular sub-problems of the analysis. To facilitate the whole investigation process, it is necessary to support interactive work, which would support the above-mentioned analytical techniques and make them easily accessible to users.

Only a few papers can be found on approaches supporting interactive
work with the data of digital evidence, which is essential for the whole forensic investigation process. Our literature survey revealed two works dealing with timelines constructed from file system activities, which are very relevant to our research.

The Zeitline~\cite{buchholz2005} tool represents activities as generic events. The user interface enables analysts to group events and then make the timeline hierarchical, to filter obtained data trees, and locate specific events by queries. 

In the CyberForensic TimeLab~\cite{olsson2009}, the timeline is implemented as a histogram using bars to represent the number of pieces of evidence at a specific time. The investigator can highlight interesting parts of the timeline and zoom in to get greater detail of that particular time span.

Both the tools are designed as generic, enabling analysts to create timelines from multiple resources, e.g., from file system metadata as well as system logs, and their user interfaces reflect this universality. In contrast, our approach focuses solely on file snapshots build from metadata only. We aim to make the analysis of this specific data maximally effective, focusing not only on the timeline but also on other data available for files. To reach this goal, we follow a user-centered design methodology, which is extended with a mechanism guiding the investigator during the process. Although our design shares some visual elements with the CyberForensic TimeLab, e.g., histograms, our solution provides an interface fine-tuned for a single specific use case -- a forensic analysis of file system snapshots. On the other hand, the visual-analytics concepts proposed in this paper are sufficiently general that they could be extended to other types of timeline in the future. 

\section{Design Methodology} \label{sec:methodology}

In this project, we applied the user-centered approach guided by the design study methodology framework~\cite{Sedlmair2012}, mainly reflecting its \textit{core} stages: discover, design, implement, deploy.

In the \emph{discover} stage, we gained a better understanding of the workflows of the digital investigation and elicited user requirements on the tool in order to simplify the analytical tasks.

The initial insight into the application domain was provided by a co-author of this paper, who is a member of the cybersecurity team of Masaryk University. Based on his initial input, we conducted semi-structured informal interviews with two other domain experts who have long-term experience with practical investigations of cybersecurity incidents. The first respondent works as a senior security specialist at CESNET -- an academic institution in the Czech Republic providing IT services to Czech academia. The second expert is a member of the incident response team at Masaryk University. All three of them have long-term experience with practical investigation of cybersecurity incidents. Each interview lasted about two hours.

Based on these interviews, we distilled a generic workflow of the investigation process and formulated requirements for a file system analysis. The results are presented in Section~\ref{sec:requirements}.

In the \emph{design} stage, we proposed the visual elements and the interactive dashboard reflecting the functional requirements. The design was proposed and refined iteratively. User interfaces were continuously prototyped under consultation with the domain expert (co-author of the paper). Proposed visual encoding is described in Section~\ref{sec:design}.

In the \emph{implement} stage, we iteratively developed the analytical dashboard. We paid attention to the observation that cybersecurity experts investigate incidents rarely, and evidence collection is a long-term interactive process. Architecture and implementation of the tool are described in Section~\ref{sec:system}.

In the \emph{deploy} stage, we evaluated the tool. 
As the investigation of real cybersecurity incidents is a sensitive process, we could not perform a usability study \emph{in the wild}. Moreover, as the developed tool deals with only part of this process, we conducted a qualitative evaluation focused directly on the tool. However, we used data from a real incident. The evaluation is described in Section~\ref{sec:evaluation} and results are summarized in Section~\ref{sec:discussion}.

\section{Requirement Analysis} \label{sec:requirements}

The interviews conducted during the \emph{discover} stage of the design methodology revealed that incident investigators would benefit from an interactive tool for file system exploration. Specific requirements were inferred from the characteristics of the data and the analytical workflow.

\subsection{Data Characteristics and Abstraction}
\label{sec:data-characteristics}

The investigation of cybersecurity incidents aims to provide
answers to key questions related to the incident, like when the activities
happened, what data was changed during the incident, where the activities
originated from, etc. The process of investigation is driven by methodologies
stipulated by digital forensics. The whole process comprises three main
stages during which the evidence is acquired, analyzed, and the final report is produced. A
simplified schema of the process is depicted in Figure~\ref{fig:workflow}.

During the acquisition phase, the investigator needs to identify and collect
the data that is likely to provide evidence about the case.
The number of possible data sources from which digital evidence can be collected is vast. In case of forensic examinations performed
directly on the machine, it is common to gather data from permanent storage (hard disk or external device like USB storage). There are also other sources of digital evidence, such as network traffic or its metadata, state and content of volatile memory, or information about authentication attempts. The rest of the paper deals with analysis of files and their metadata. It keeps the investigation domain limited in size while making it possible to evaluate the main principles.

File metadata describes information about the file, maintained by the operating system together with the file data. The exact scope of metadata depends on the operating system used, however, nowadays, it is common for all widely used file systems to recognize the file name, file ownership (specifying the user and a group), content size, and access rights.
Besides these, several timestamps are maintained, indicating the time when key activities with the file or the metadata were last performed:

% is a common set of attributes that are supported by all widely used filesystems nowadays:

\begin{itemize}[itemsep=0mm]
\item \emph{a-time}: the time when the file content was last read (accessed),
\item \emph{m-time}: the time when the file content was last modified,
\item \emph{c-time}: the time when the metadata record was last changed (e.g., during the change of access rights),
\item \emph{b-time}: the time when the file was created. The \emph{b-time} timestamp is supported only by advanced file systems.
\end{itemize}

All the timestamps, except for \emph{b-time}, change during the file life-time based on the operations performed. When a timestamp is updated, the previous value is overwritten and lost, which means they always refer only to the last performed actions.

%okomentovat timestampy (zmeny souboru znamena zmenu metadat -- velikost), nektere nelze menit

Timestamps are an essential source of information for the reconstruction of events relevant to the investigation. They can help understand when certain operations took place but also reveal the nature of the activities performed. For instance, when a file is copied from another computer, the copying process usually retains the original timestamp. Such a file has the \emph{m-time} value set to a date before the \emph{b-time} and \emph{c-time} values, which both will refer to the time when the copying process finished. A brand-new file created on the system has all the timestamps set to the same value upon creation. The difference in the timestamps can reveal where the file originates from.

Even if they do not reveal the actual file content, all file metadata attributes play a big role in the incident analysis. One of the most important reconstructions is determination of the timeline of actions performed in the analyzed system. A timeline emphasizes crucial activities conducted during the incident. For instance, it specifies when the attacker accessed the system for the first time or when a specific system configuration got changed.

A timeline constructed from metadata is a list of records ordered by the timestamps. Since there are multiple timestamp types assigned to a file, a single file can occur multiple times in the list, whenever its timestamps differ. A typical timeline contains hundreds of thousands of records, which need to be further analyzed.

In addition to providing input to recover the timeline, metadata can be used for efficient filtering of files, based on unique \emph{fingerprints} they form, such as similarities of file locations, common access rights, or suspicious ownership.

\subsection{Requirements}

Based on the interviews, data abstraction, and the analytical workflow, we identified five functional requirements:

\textbf{R1: Exploration of the file system structure.} 
%During the investigation, the analysts have to pay attention to different parts of the file system, e.g., files in a specific directory, files with a specific extensions, all log files, all authorization files, etc. Moreover, they often switch between different subjects of interest. Therefore, the analytical tool should support analysts in the efficient navigation and filtering based on the location of files and directories, their meaning, utilization, and other semantics encoded in the attributes. 
During the investigation, the analysts have to pay attention to different parts of the file system, e.g., files in a specific directory, files with specific extensions, or all log files. However, the interviewed domain experts emphasized that the interactive hierarchical exploration of the file system is not helpful. Instead, they need a global temporal view of the file system data with the possibility to navigate in the file system structure effectively. The analytical tool should support analysts in the efficient switching between different parts of the file system and narrowing the area of interest by offering filtering functions that would localize the data by various aspects and meaning encoded in the available file system metadata.
% Design:
%  - clustering, keyword search, pins in the histogram, sorting and filtering functions in timeline view
% Evaluation tasks:
%  - all (generic requirement)

% Jestli se to vztahuje k stromovemu zobrazeni adresaru/souboru (ala napr. Windows Explorer), tak to je neco, co nas z pohledu analyzy zajima minimalne. My potrebujeme resit primarne linearni seznam (timeline) a resit jeho redukci. Kontext, v jakem jsou soubory ulozeny neni zajimavy pro ty analyzy, ktere Fimetis resi.

\textbf{R2: Exploration of temporal relationships.} 
Disk snapshots have strong temporal characteristics. Each record provides the timestamp of the last manipulation, e.g., the creation, modification, or access. However, every file or directory usually appears multiple times in the dataset as the manipulation timestamps differ, which increases the data volume to be inspected. Also, the recorded data period is often very long, containing timestamps from a time long before the system was installed (but from when the files were created). Therefore, providing a scalable temporal view on the data with efficient filtering, zooming, and preserving time coherence is very important for making the analysis effective.
% Design:
%  - interactive time-span manipulation in histogram, automated aggregation of the data in histogram with details on demand, ...
% Evaluation tasks:
%  - all (generic requirement)

\textbf{R3: Detection of file system anomalies.}
Some combinations of file locations and attributes can be considered unusual or deserving analyst's attention. For example, publicly writable files or directories, hidden files outside of users' homes, executables with administrator's privileges, files masking their names (e.g., a binary file with a \emph{.txt} extension or named with only white spaces). The analytical tool should provide multiple views on various combinations of location paths and attributes in order to localize potential anomalies easily, and then further explore the corresponding files using \textbf{R1} and \textbf{R2} principles.
% Design:
%  - predefined clusters (user-defined clusters in the future work)
% Evaluation tasks:
%  - T1: Files or directories with suspicious names
%  - T2: System files (configurations or executables) modified by the attacker
%  - T4: Privileged executables (with root s-bit) used in the attack
%  - T6: Possibly compromised users (changes in the configuration of user accounts)

\textbf{R4: Traces of the execution of suspicious commands.} % was R3
Some commands are seldom used by administrators but often used by attackers. For example, the \texttt{shred} Unix command is often used to wipe data content. The tool should allow analysts to verify whether or not such commands were used. Command execution can be identified by the \emph{a-time} attribute. Once the command execution is confirmed, the analyst can use interactions reflecting \textbf{R1} and \textbf{R2} to explore details, analyze the impact of the execution, and either confirm or reject the hypothesis that an attacker executed the command.
% Design:
%  - a specialized cluster
% Evaluation tasks:
%  - T5: Suspicious (unusual) commands possibly executed by the attacker

\textbf{R5: Traces of batch processing.} % was R4, R5
Besides the execution of specific commands (\textbf{R4}), attackers often use scripts to perform reconnaissance on the system or to compile programs or libraries before installing them into the system. These batch activities can be recognized by the execution of multiple commands or the creation of multiple files in a short time, while manual tasks take a longer time. However, batch processing can represent a legal activity, e.g., the legal compilation or the result of regular system updates. Therefore, the tool should support analysts in efficiently identifying batch processes in the huge amount of file system data and then allowing them to analyze suspicious activities further using \textbf{R1} and \textbf{R2}.
% Design:
%  - a specialized cluster for suspicious commands and signs of compilation
%  - na histogramu obecne muzeme videt anomalie ve zmenach a pouzivani FS, coz muze byt nekdy indikator batch procesu. Jinak primo podporujeme detekci kompilace (viz cluster) a naslednou analyzu pomoci R1 a R2.
%  - Future work: Select a dynamic filter enabling the analyst to set time span for batch processes, time span for manual processes, or their ratio. The analyst should see the impact of the filter parameters immediately (e.g. number of identified batch processes)
%  - T3: Executables or libraries that were NOT installed from its package, i.e., downloaded or manually compiled

While the requirements \textbf{R1} and \textbf{R2} reflect the generic investigation workflow, requirements \textbf{R3--R5} are related to more specific analytical questions that are often asked during the file system investigation. Besides these functional requirements, we set two complementary qualitative requirements that affect the architecture and implementation. These requirements follow the practice emphasized by the interviewees where cybersecurity experts investigate incidents rarely, and every investigation takes a lot of time (hours or days). 

\textbf{R6: Easy to use.} Even practicing incident investigators analyze disks rarely (see Section~\ref{sec:evaluation}). Therefore, they should be able to use the tool even after a long period without the need for repeated learning. 

\textbf{R7: Persistence.} The data and interactions have to be persistent so that an analyst can pause the investigation process and continue later on. Persistence is also important for recalling previous investigations and comparing hypotheses and results.

%\textbf{R8: Availability.} For a quick response, the tool should be available at any time without the need to install anything locally. 

\section{Visual Design} \label{sec:design}

In this section, we summarize the design rationale, visual encoding, and interaction capabilities.
The user interface consists of three coordinated views~\cite{roberts2007,scherr2008}, where a change in one view to the dataset affects other parts of the dashboard. 

\subsection{List View} \label{sec:list-view}

The \emph{List View} (\autoref{fig:teaser} -- A) is a dominant part of the dashboard providing a view on the raw data. Records are sorted by the timestamp by default (\textbf{R2}), but they can be re-ordered according to the file system structure (\textbf{R1}) by clicking on the \emph{File Name} or \emph{Type} columns. Individual columns can be shown or hidden via the \emph{List View} menu (the three dots in the up-right corner of the \emph{list view} area).

\begin{figure}[!htbp]
  \centering
  \includegraphics[width=.98\columnwidth]{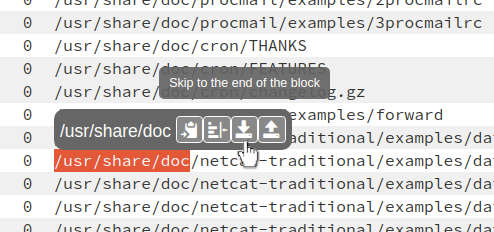}
  \caption{Detail of smart block skipping in the \emph{List View}.}
  \label{fig:smart-skip}
\end{figure}

Analysts can browse records traditionally by scrolling the list up and down, or they can use \emph{smart block skipping} (\autoref{fig:smart-skip}) that significantly increases the efficiency of the list exploration. By clicking on a timestamp or a file path, the prefix is highlighted, and a context menu appears that enables analysts to skip records with the same prefix. Using this feature, analysts can quickly navigate to the next or previous date, hour, or sub-directory, and then accelerate the data exploration either from structural (\textbf{R1}) or temporal (\textbf{R2}) perspective.

The background of lines with the same timestamp is brushed to visually distinguish different time blocks (\textbf{R2}).

Search operation in the list works at two levels (the \emph{name selection} label in~\autoref{fig:timeline-histogram}). Typing text into the input search field highlights the corresponding parts of the file paths. If the text is confirmed or the user clicks at the magnifier icon, then the list of records is filtered out, and only relevant lines remain displayed, enabling the analyst to pay attention to only desired files and directories (\textbf{R1},\textbf{R4}). Data filtered out in this way remains in the \emph{Histogram} (see~\autoref{sec:histogram}) to preserve a broader context, but they are grayed out.

Records of high importance can be bookmarked (the \emph{bookmarks} label in~\autoref{fig:timeline-histogram}). Bookmarked records are emphasized in the list, displayed in the \emph{Histogram} view, and used for fast navigation (\textbf{R2}). Bookmarks are persistent throughout the whole analysis and can be removed only on demand. Moreover, as they provide a broader context with significant events, the bookmarked lines are always visible in the \emph{List View}, even if they do not fit all filters of the dashboard at the moment.

\subsection{Histogram} \label{sec:histogram}

The \emph{Histogram} section (\autoref{fig:teaser} -- B) provides an interactive view on data distribution. 

\begin{figure*}[!htbp]
  \centering
  \includegraphics[width=.98\textwidth]{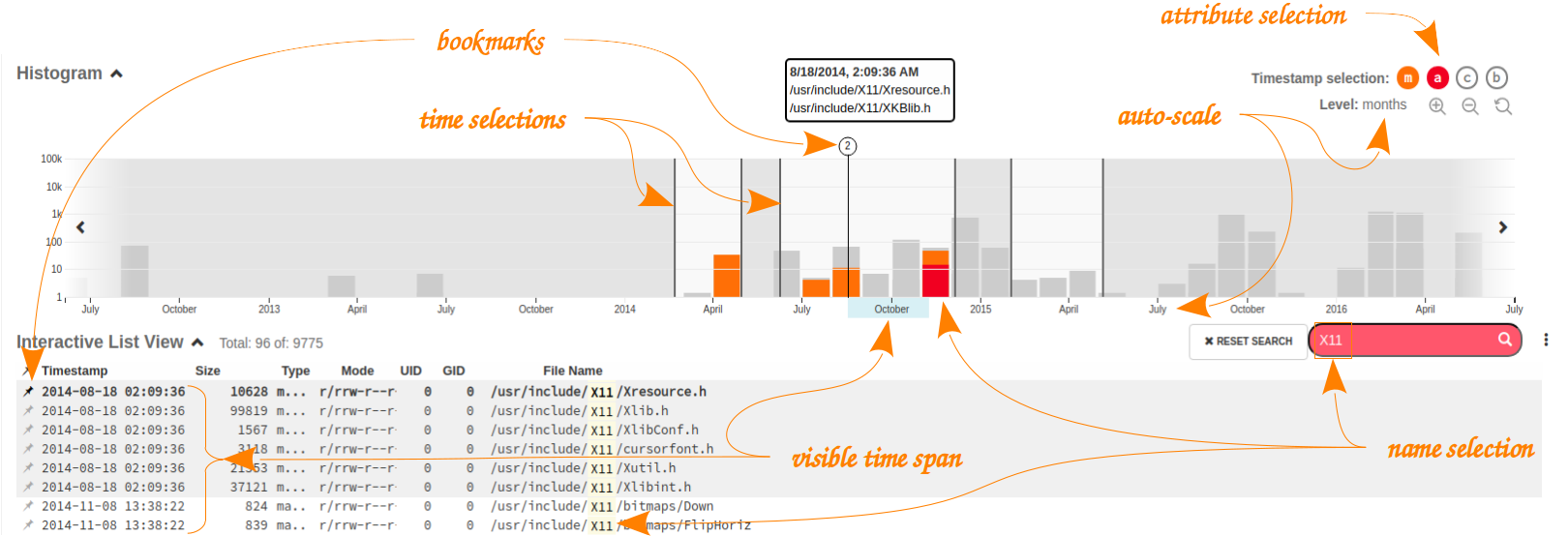}
  \caption{Navigation and filtering in the \emph{List View} and \emph{Histogram}.}
  \label{fig:timeline-histogram}
\end{figure*}

The \emph{y}-axis encodes the number of records. The axis has a logarithmic scale to deal with high peaks that often appear in the data but still preserve the visibility of low numbers that can be important for analysts.  

The \emph{x}-axis is scaled automatically (the \emph{auto-scale} label in~\autoref{fig:timeline-histogram}). When zooming in, the \emph{x}-axis automatically changes from years to months, days, and hours, and vice versa. The bars are recalculated and aggregated accordingly, representing the distribution in a specific year, month, day, etc. Zooming can be performed either by mouse, keyboard, or via icons in the upper-right corner.

Different colors in the histogram encode different file system operations (values of the \emph{Type} column in the \emph{List View}). Color encoding is shown in the \emph{Timestamp selection} section. A detailed description of the metadata attributes is provided when the mouse is located over an icon. Similarly, hovering the mouse pointer above a bar in the histogram triggers a pop-up tool-tip with attribute type, time, and an exact number of records. Clicking on a bar scrolls the \emph{List View} to the corresponding entries.

The \emph{Timestamp selection} is also used for per-attribute filtering (the \emph{attribute selection} label in~\autoref{fig:timeline-histogram}). Attributes can be switched on or off in the histogram by clicking on the icons. The \emph{List View} is updated accordingly -- only the records with selected attributes are shown in the list. 

The histogram also serves as a time focusing tool (the \emph{time selection} label in~\autoref{fig:timeline-histogram}). Using a mouse, the analyst can draw multiple span windows and thus restrict the lines shown in the \emph{List View}. A context menu appears when a user selects a selection span window. This menu enables the user to perform common operations, like extending the span, zooming into the span, or erasing the span. Some of these operations are available via direct mouse interaction in the histogram as well.

Due to restricted space on the web page, the \emph{List View} displays only part of all the records at any one time (the rest is available via scrolling). Visible records represent span, which is emphasized in the \emph{x}-axis of the histogram as a cyan stripe (the \emph{visible time span} label in~\autoref{fig:timeline-histogram}). This stripe supports the visual correlation between the \emph{List View} and the histogram.

Entries bookmarked in the \emph{List View} are shown in the histogram as push-pin icons. If they are too dense, they are aggregated into a single icon with a number of merged bookmarks. Details are provided as a tool-tip triggered on the mouse hover. Click on the icon scrolls the \emph{List View} into the corresponding entry (to the first record in the case of aggregated push-pin). Push-pins that are out of selection spans are not clickable.

Span selectors, bookmarks, and automatically adaptable \emph{x}-axis represent a powerful combination enabling analysts to scale and explore data from the time perspective (\textbf{R2}). 

The structural exploration (\textbf{R1}) is less dominant in the histogram view. It is mainly restricted to the per-attribute filtering of records. On the other hand, the per-attribute filtering combined with the path filtering of the \emph{List View} provides a generic approach to solve \textbf{R3} and \textbf{R5}. For example, a C/C++ compilation process accesses header files and the \texttt{gcc} compiler binary. A proper combination of the filters can reveal these traces. Moreover, the compilation unusually touches a huge amount of header files, leaving peaks in the histogram, especially when performed in calm nighttime. 

\subsection{Clusters} \label{sec:clusters}

\emph{Clusters} (\autoref{fig:teaser} -- C) represent a generic mechanism enabling analysts to select files or directories with a specific "fingerprint". Clusters are defined by the combination of modification attributes (entries with \emph{m-a-c-b} modification types) and regular expressions applied to the file names. Taking into account analytical requirements \textbf{R3 -- R5} and needs of domain experts, we predefined several clusters covering the most common investigation tasks for UNIX file systems. Additional clusters can be easily appended.
\begin{compactitem}
    \item \emph{All files} --The default cluster with no filtering.
    \item \emph{User SSH files} -- Configuration files and SSH keys stored in the users' home directories.
    \item \emph{Standard executables} -- Files stored in the standard system directories for binaries, e.g., \texttt{/bin}, \texttt{/sbin}.
    \item \emph{Python/shell/PHP/perl scripts} -- Several clusters based on standard file extensions, e.g. \texttt{.py}, \texttt{.sh}.
    \item \emph{Cron definitions} -- Files stored in the default locations of \texttt{cron} jobs, i.e., regularly executed services.
    \item \emph{Starts with '.'} -- Hidden files or directories.
    \item \emph{Suspicious files} -- Files or directories with names consisting of dots and white spaces.
    \item \emph{Executables with sbit} -- Executables that can run under a different user or group privileges than the original user or group.
    \item \emph{Weak permissions} -- Executable files writable for general users.
    \item \emph{Compilation signs} -- Access to C/C++ header files and the compiler executables.
    \item \emph{Unusual commands} -- Commands that are rarely used by common system administration, but often by attackers, e.g., \texttt{wget}, \texttt{curl}, and \texttt{shred}.
    \item \emph{System configuration changes} -- Important files related to the system configuration, e.g., \texttt{/etc/init.d} or \texttt{/etc/passwd}.
\end{compactitem}

In the current implementation, only one cluster can be selected at one time. The number of all records fulfilling cluster criteria is shown as a ``total entries'' number. The ``filtered entries'' indicator shows the number of records satisfying other filtering criteria of the dashboard, and then they are listed in the \emph{List view} and included in the \emph{histogram}. A bar under each cluster box visually emphasizes the ratio between the filtered and total records, enabling the analysts to identify the impact of currently used filtering criteria on clusters.

%Clusters can be shown/hidden in/from the cluster area via menu.

\section{System Architecture and Implementation} \label{sec:system}

FIMETIS is designed as a client-server application. The client part is implemented as a web application built on the Angular framework. Interactive visualizations use the D3.js library. The server part provides services for file system data management (import, export) and interactive data processing via the client. The Flask REST API handles the client-server communication. Flask is a lightweight web server gateway interface written in Python, which mediates access to the backend API -- the center of the application logic and communication with databases. This architecture enables a concurrent investigation of multiple sources. It is possible to open two file systems simultaneously in two different explorer windows, for instance, and explore them side by side.

Persistence (\textbf{R7}) is guaranteed by two database systems. The file system snapshots are stored in the NoSQL Elasticsearch database. Configuration data, user accounts, interactions (e.g., bookmarks), and other operational data related to the analysis are stored in the rational Postgresql database.

\section{Evaluation}\label{sec:evaluation}

To gather feedback on how well the tool fulfill the requirements \textbf{R1--R5}, and to identify possible refinements for the future design process iteration, we conducted a qualitative evaluation. The evaluation was held in June 2020. 

\begin{figure*}[!htbp]
  \centering
  \includegraphics[width=.98\textwidth]{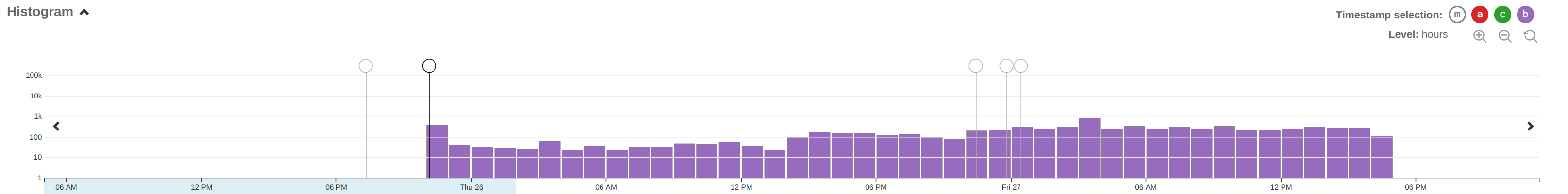}
  \caption{Indication of a continuous creation of files generated by the network scanner.}
  \label{fig:scan}
\end{figure*}

\subsection{Participants}

We conducted the user study with five cybersecurity professionals who represent the target audience of the tool. All of them are members of the university cybersecurity research team or a security team in another organization. One participant works as an incident investigator in a private company. The average age of all participants was 30.2 years (\textit{SD=3.5}); all of them were males. Two of them participated in initial interviews from which the requirements were derived. However, they did not participate on the design of the tool.

All the participants were cybersecurity professionals. However, they differ in the experience with practical investigation of incidents using file system analysis. Their skills are summarized in~\autoref{tab:demography}. 

\begin{table}[!htbp]
    \centering
    \begin{tabular}{cclcc}
        \textbf{ID} & \textbf{Age} & \textbf{Occupation} & \textbf{INC} \\
         P1 & 34 & researcher in cybersecurity            & $<$3 \\
         P2 & 32 & researcher in cybersecurity            & 0 \\
         P3 & 32 & incident investigator -- network analyst & $<$3 \\
         P4 & 26 & lead security analyst                  & $>$10 \\
         P5 & 27 & incident investigator                  & $>$10 \\
    \end{tabular}
    \caption{Demographic information of our participants. Occupation -- position related to network administration and incident investigation, INC -- number of incidents investigated by the analyst using disk analysis.}
    \label{tab:demography}
\end{table}

\subsection{Data sets}

During the evaluation, we used two datasets that were captured from computers affected by real incidents. The files were maintained using the \emph{ext4} file system, which is commonly used on UNIX servers. We used different mechanisms to capture the primary
data, yielding some records without the \emph{b-time} timestamp (see
~\ref{sec:data-characteristics}). The first dataset contained $308\,311$
records and was used for the tool demonstration and familiarization of
participants with the dashboard. The second dataset consisted of $505\,742$
records and was used for the evaluation.

We carefully analyzed the second dataset using FIMETIS to reconstruct the
incident to establish a baseline for the evaluation. Navigating through
the predefined clusters, we gradually collected a list of crucial findings relevant to the incident.
We identified six clusters that are most relevant to providing evidence of the incident. 

\begin{compactitem}
    \item \emph{User SSH files} -- Displays access to SSH key files used by the attacker to control remote access to user's account.
    \item \emph{Suspicious files} -- A bunch of files is visible in \texttt{/var/tmp/...}. The directory name is suspicious (\texttt{...} is often seen during attacks) and it contained files named using IP addresses, suggesting it was used as a cache for network scans.
    \item \emph{Executables with sbit} -- In addition to standard Unix commands, the output reveals file \texttt{/var/lib/.s}, which is definitely not legit (tries to hide itself and elevates the executable rights using the root s-bit parameter).
    \item \emph{Unusual commands} -- Two HTTP command-line clients can be seen in the output that are used recently: \texttt{wget} and \texttt{curl}.
    \item \emph{System configuration changes} -- Changes to the machine user accounts can be identified in the output.
    \item \emph{Compilation signs} -- Several compilations of C-language codes are present in the dataset.
\end{compactitem}

However, these pieces of evidence are often hidden in a huge amount of other entries. Therefore, using the list view and histogram is necessary to focus attention on relevant parts of the dataset. Having put all the collected information together, we compiled a precise summary of the incident and its timeline:

\begin{enumerate}[itemsep=0mm,label={S\arabic*:}]
    \item 2016-05-25, 00:40: The attacker illegally logged in the account of user \emph{martin} using SSH for remote access. Further analysis showed that the attacker abused unsecured NFS access to \texttt{/home} directory, allowing to upload of files and execution of privileged binaries. This is the only part of the analysis that could not be done just with the file system metadata, but the provided file system evidence gave a precise lead about what to check in the system logs and configuration.

    \item 2016-05-25, 02:40: The attacker installed a trojan code. A purportedly malicious \texttt{libselinux} library was downloaded using the \texttt{wget} command, and the system configuration (in file \texttt{/etc/ld.so.preload}) was changed to likely inject the library into every newly created process. The SSH service was restarted to activate the trojan code (either a backdoor and/or credential-stealing). A suspicious s-bit file \texttt{/var/lib/.s} was installed simultaneously, probably to trigger the illicit activities.
    
    \item 2016-05-25, 19:20: There are suspicious activities in the account of user \texttt{roberto}. This account was probably also compromised a few hours later by the attacker as both the accounts show similar signs, e.g., an empty file named \texttt{1}. The reason is uncertain. However, there is no evidence that this account was used for suspicious activities.
    
    \item 2016-05-25, 21:22: The attacker re-compiled and re-installed the trojan code. The attacker was probably not satisfied with the version they deployed at the beginning of the day, so they returned, re-compiled the \texttt{libselinux} library, and then produced another binary on the spot.

    \item 2016-05-25, 22:08: The attacker created a hidden directory `\texttt{/var/tmp/...}`, where they compiled some suspicious tools, e.g., \texttt{pcap} or \texttt{nmap}, and installed them into the system. Following that, they started a network scan and used the directory to store results obtained for individual network targets. Since then, the data was kept being captured and logged into this directory. The directory is used for a massive scan spanning almost two days, which is visible from the relevant histogram, see~\autoref{fig:scan}.

    \item 2016-05-26, 23:12: The system files with user account and passwords (\texttt{/etc/shadow} and \texttt{/etc/passwd}) were modified one day later. It is uncertain whether this activity is related to the incident or not.
\end{enumerate}

\subsection{Apparatus}

The server part of the FIMETIS application was deployed on a common cloud machine, equipped with 8GB RAM, 80GB disk space and 4 CPUs.
We conducted the evaluation online using Google Meet. The participants used Google Chrome on their computers or laptops with resolutions ranging from FullHD to UHD. Their interaction and comments were recorded for later analysis. 

\subsection{Procedure}

The user study was divided into four parts. First, the participants were introduced to the general procedure, signed a consent form, and filled the demography questionnaire. Then, the experimenters presented the tool, explained all its features using the first dataset, and let the participant familiarize with the tool for 5--10 minutes.

Next, the participants were to find the following signs of the file system manipulation and usage:
\begin{enumerate}[itemsep=0mm,label={T\arabic*:}]
    \item Files or directories with suspicious names.
    \item System files (configurations or executables) possibly modified by the attacker.
    \item Executables or libraries that were not installed from its package (i.e., either directly downloaded or manually compiled on the system).
    \item Privileged executables (with root s-bit) possibly used in the attack.
    \item Suspicious or unusual commands possibly executed by the attacker.
    \item Possibly compromised user accounts.
\end{enumerate}

These tasks address requirements \textbf{R1--R5}. Together, they should provide an overview of what happened during the incident. While the tasks \emph{T1,T2,T4, and T6} reflect different aspects of the detection of file system anomalies (\textbf{R3}), \emph{T5} and \emph{T3} are related to the execution of suspicious commands (\textbf{R4}) and traces of batch processing (\textbf{R5}) respectively. All the tasks require iterative exploration of the file system structure (\textbf{R1}) and temporal relationships (\textbf{R2}). 

The participants had the tasks printed out so that they could easily make notes. The experimenter asked the participants to solve the tasks iteratively in any order. They were asked to think aloud. At the end of this evaluation phase, they had to summarize the incident upon their observations.

Although the real investigation of an incident lasts many hours or can even spread to several days, we restricted the participants to roughly one hour. The study's goal was not to get all the details about the attack, which is usually not possible without additional pieces of information such as system logs or network traffic, but to ascertain whether the analyst can get a quick insight into the incident using our tool.

When the incident investigation ended, the participant filled the usability questionnaire (Simple Ease Question, SEQ~\cite{Sauro2009}), and System Usability Scale, SUS~\cite{Sauro2011}. Finally, the experimenter interviewed participants on their final thoughts and feature requests.

\subsection{Limitations}

This user study has several limitations. The number of participants is relatively low. The reason lies in the time demands put on the evaluation process, which took roughly two hours per participant. To minimize the impact of this limitation, we involved security practitioners -- possible users of the tool. On the other hand, we aimed to cover a wide range of expertise. Therefore, we engaged both highly skilled experts who have practical experience with collecting evidence from file systems and professionals who lack these specific skills as they focus on other cybersecurity domain, e.g., network analysis or cybersecurity research.

We are also aware that the evaluation was performed with only one test case, and then the results could be affected by the specific attack vector hidden in the dataset. We strove for authenticity, and then we preferred a real incident from artificial data. On the other hand, we aimed to choose an incident which is typical in a sense. The selected dataset contains the digital evidence of common attack steps like the abuse of user accounts, privilege escalation, installation of backdoor, and using the compromised host for further illegal activities.

\subsection{Results}

\paragraph{Usability \& learnability:} User experience with the tool was evaluated by the System Usability Scale (SUS). SUS is a de facto standard method for assessing systems' usability regardless of their purpose. The average SUS score of FIMETIS was 88.5. According to the adjective ratings~\cite{Bangor2009}, the score corresponds to \emph{excellent} ratings and proves compliance with \textbf{R6}. 

SUS questions \#4 (I think that I would need the support of a technical person to be able to use this product) and \#10 (I needed to learn many things before I could get going with this product) can also be used to interpret learnability~\cite{Sauro2011}. The average answers 1.2 and 1.8, respectively, on the Likert scale from 1: \enquote{strongly agree} to 5: \enquote{strongly disagree} suggest that FIMETIS is also easy to learn.

\paragraph{Preferences in using visual-analytic elements:} FIMETIS is designed as a generic tool where hypotheses can be verified in various ways using the combination of diverse visual-analytical elements. To explore if some elements are more popular then other, we analyzed videos captured during the evaluation. We measured the usage of key interactions and data filtering concepts: filtering data by attributes, using predefined clusters, filtering data by span windows, searching and filtering by path, and using push-pins. 

The results are summarized in~\autoref{fig:radar-chart}. Push-pins represent the maximal number of bookmarks used by the analyst at the same time (20 push-pins in the participant P5). The other axes encode the relative time the analyst used the element. The time is expressed as the percentage of the investigation time. It is to be pointed out that the \emph{name filtering} is used occasionally for temporal filtering and navigation during the interaction with the \emph{List View}. Therefore, its usage can be underestimated in the radar charts.

\begin{figure*}[ht]
  \centering
  \includegraphics[width=.8\textwidth]{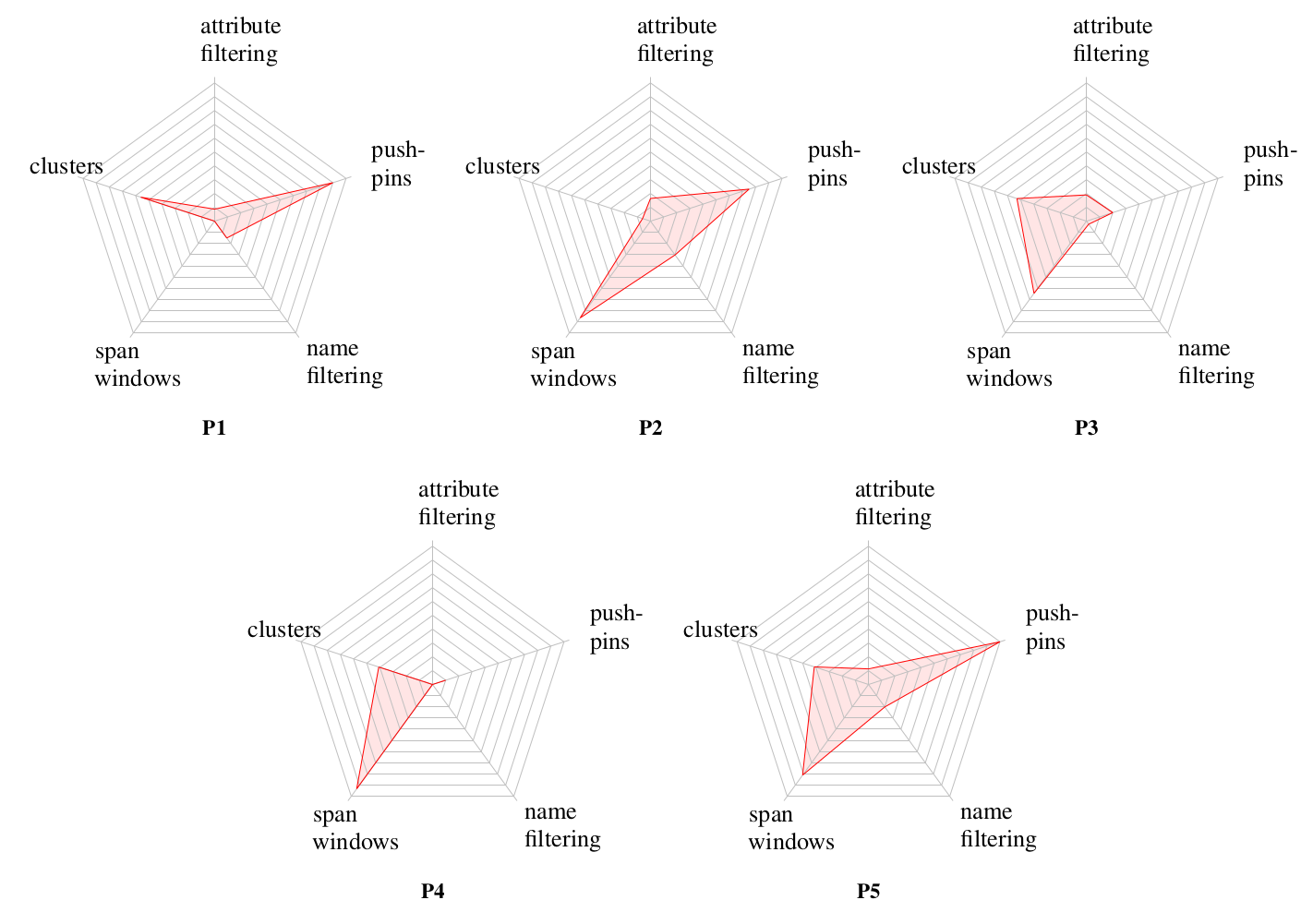}
    \caption{Approximate utilization of visual-analytic elements of GUI by individual participants P1--P5. The \emph{push-pins} axis encodes maximal number of bookmarks used simultaneously. Other axes represent the relative time (as the percentage of investigation time) when the element was used.}
  \label{fig:radar-chart}
\end{figure*}

The radar charts depicted show that different analysts preferred different combinations of elements. Usually, only 2--3 elements are used intensively, while others are ignored either completely or used significantly less. Another interesting observation, which is not captured in the radar charts, is that the analysts used only one span window. P1 did not use this element, and P3 used two span windows simultaneously, but only for a very short time.

\paragraph{Precision of the attack timeline:} To evaluate the ability of the FIMETIS tool to provide a quick insight into the incident timeline, incident scenarios reported by participants were compared with the baseline scenario \emph{S1--S6}. The precision was ranked by the authors of the paper. The results are summarized in table~\ref{tab:precision}.

\begin{table}[!htbp]
    \centering
    \begin{tabular}{ccccccc}
        & \textbf{S1} & \textbf{S2} & \textbf{S3} & \textbf{S4} & \textbf{S5} & \textbf{S6} \\
        \textbf{P1} & \fullpie & \halfpie & \fullpie & \emptypie & \fullpie & \fullpie \\
        \textbf{P2} & \fullpie & \fullpie & \fullpie & \emptypie & \fullpie & \fullpie \\
        \textbf{P3} & \halfpie & \halfpie & \halfpie & \emptypie & \fullpie & \fullpie \\
        \textbf{P4} & \fullpie & \fullpie & \fullpie & \halfpie  & \fullpie & \fullpie \\
        \textbf{P5} & \halfpie & \halfpie & \halfpie & \emptypie & \fullpie & \fullpie \\
    \end{tabular}
    \caption{Precision of the attack reconstruction: \emptypie[.9ex] overlooked/not identified, \halfpie[.9ex] identified partially, \fullpie[.9ex] identified correctly.}
    \label{tab:precision}
\end{table}

\emph{S1 (compromising the account 'martin')} was identified by all participants. However, P3 and P5 identified the account together with 'roberto'. They did not decide who was the primary target of the attacker.

\emph{S2 (installation of a trojan code)} was identified by all participants, but the level of observed details varied. All the participants discovered the \texttt{/var/lib/.s} as part of the attack vector, but P1, P3, and P5 did not provide more details about this attack phase. Moreover, the \texttt{selinux} library was completely overlooked by them. P2 did not mention the restart of the SSH server, but SSH was correctly identified as the service used for the escalation of privileges. P4 noticed and described all the details related to this attack phase, including the usage of \texttt{/etc/ld.so.preload}.

\emph{S3 (suspicious manipulation with the account 'roberto')} was identified by all participants and considered part of the attack. Neither participant found the real abuse of this account. However, P3 and P5 did not decide whether the 'roberto' or 'martin' was the primary access point for the attacker.

\emph{S4 (re-compilation and new installation of the trojan code)} was overlooked by all participants except P4. This analyst noticed the re-installation but overlooked the re-compilation of the trojan code at the compromised computer.

\emph{S5 (a hidden directory)} was identified by all participants very quickly. The directory contained almost 12.000 records combining source code of multiple tools, traces of their compilation and usage, and data files gathered by the attacker. Nevertheless, the analysts were able to spot tools and data relevant to the attack vector and directly describe their purpose in the attack (P2, P3, P4, P5) or at least mention them as a tool worth further exploration (P1).

\emph{S6 (modification of the user account database)} was identified by all participants. P1 noticed the changes but finally considered as not being linked to the incident. P2 did not provide more details. Other analysts considered the changes to be part of the attack when the attacker probably created a new user for later access. 

\begin{figure}[ht]
  \centering
  \begin{tikzpicture}[scale=1.0]
  \begin{axis}
    [
      boxplot/draw direction=x,
      ytick={1,2,3,4,5,6,7}, % sloty
      yticklabels={T6, T5, T4, T3, T2, T1}, % nazvy
      xtick={1,2,3,4,5},
      x axis line style = {opacity=0},
      ytick style = {draw=none},
      xtick pos=left,
      ytick pos=left,
      height=5cm,
      xmin=0.8, 
      xmax=5,
      boxplot/every median/.style={black,very thick,solid},
      cycle list={}, % to avoid dashed lines when we have more than 5 tasks
    ]
    \addplot+[ % T6
    	color=black,
    	boxplot prepared={
        	box extend=0.5, whisker extend=0.3,
        	%median=2, 
        	upper quartile=2, lower quartile=2, upper whisker=4, lower whisker=1, average=2.2
        },
    ] coordinates {};
    \addplot+[ % T5
    	color=black,
    	boxplot prepared={
        	box extend=0.5, whisker extend=0.3,
        	%median=2, 
        	upper quartile=2, lower quartile=1, upper whisker=2, lower whisker=1, average=1.6
        },
    ] coordinates {};
    \addplot+[ % T4
    	color=black,
    	boxplot prepared={
        	box extend=0.5, whisker extend=0.3,
        	%median=2, 
        	upper quartile=2, lower quartile=1, upper whisker=4, lower whisker=1, average=2
        },
    ] coordinates {};
    \addplot+[ % T3
    	color=black,
    	boxplot prepared={
        	box extend=0.5, whisker extend=0.3,
        	%median=5, 
        	upper quartile=5, lower quartile=3, upper whisker=5, lower whisker=2, average=3.8
        },
    ] coordinates {};
    \addplot+[ % T2
    	color=black,
    	boxplot prepared={
        	box extend=0.5, whisker extend=0.3,
        	%median=2, 
        	upper quartile=3, lower quartile=2, upper whisker=4, lower whisker=2, average = 2.6
        },
    ] coordinates {};
    \addplot+[ % T1
    	color=black,
    	boxplot prepared={
        	box extend=0.5, whisker extend=0.3,
        	%median=1, 
        	upper quartile=2, lower quartile=1, upper whisker=2, lower whisker=1, average=1.4
        },
    ] coordinates {};
  \end{axis}
\end{tikzpicture}
  \caption{Distribution of answers to SEQ tasks (min/max values, lower/upper quartile, and average). Lower score is better (1 = Very easy, 5 = Very difficult).}
  \label{fig:boxplot}
\end{figure}
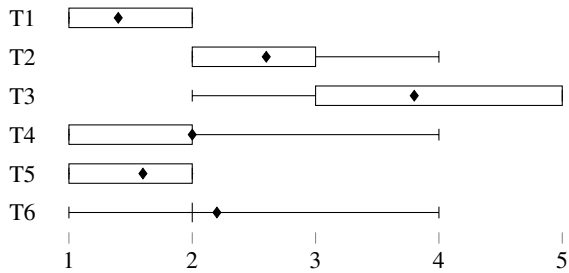

\paragraph{Tasks difficulty:} To evaluate the usability of the tool for solving individual tasks \emph{T1--T6}, we analyzed the SEQ answers. We used this method because our tasks were too complex for metrics such as task duration time or completion rate, and the method performs as good as more complicated measures of task difficulty~\cite{Sauro2009}. The participants responded to a single question associated with individual tasks (\enquote{Overall, how difficult or easy did you find this task?}), using a scale from 1 (very easy) to 5 (very difficult). The box plot is depicted in~\autoref{fig:boxplot}. 

Overall, the participants considered tasks rather easy with the FIMETIS tool. This result correlates with the analysts' success to correctly reconstruct the incident in limited time at an appropriate level of detail. The only exception was finding out executables or libraries that were not installed from its package (\emph{T3}). This task is considered rather difficult. However, this result also corresponds to the low success rate of revealing the re-compilation of a trojan code (step \emph{S4} of the incident). The reason probably lies in the complexity of the task, which forces the analyst to iteratively combine multiple views and combine multiple features of the tool. 

\iffalse
\begin{figure}[ht]
  \centering
    \begin{tikzpicture}[scale=0.51, rotate=18]
    \tkzKiviatDiagram[scale=0.51, label space=3, gap=1, lattice=10, radial  style/.style={gray!50,-,shorten >=0.1cm}] {
        push-pins,
        attribute\\ filtering,
        time\\ windows,
        clusters,
        name\\ filering
    }
    \tkzKiviatLine[thick,color=blue,mark=none,fill=blue!20,opacity=.5](9.0,0.866,0.000,5.612,1.507)
    \tkzKiviatLine[thick,color=green,mark=none,fill=green!20,opacity=.5](7.5,1.643,8.651,0.604,3.023)
    \tkzKiviatLine[thick,color=red,mark=none,fill=red!20,opacity=.5](2.0,1.897,6.455,5.288,0.275)
    \tkzKiviatLine[thick,color=orange,mark=none,fill=orange!20,opacity=.5](1.0,0.000,9.333,4.087,0.000)
    \tkzKiviatLine[thick,color=gray,mark=none,fill=gray!20,opacity=.5](10,1.129,8.105,4.122,2.008)
    %\tkzKiviatGrad[prefix=,unity=10,suffix=\%](1)  
    \end{tikzpicture}
  \caption{TO DO.}
  \label{fig:radar-chart}
\end{figure}
\fi

\section{Discussion and Future Work} \label{sec:discussion}

The work we presented in this paper focuses on the design and user evaluation of a visual-analytics tool that aims to support efficient disk snapshot exploration as part of the cybersecurity incident investigation workflow.

We collaborated with three skilled investigators on the clarification of forensic processes and the specification of requirements. The evaluation conducted with five cybersecurity experts revealed that the analytical tool built upon these requirements is intuitive and easy to use. All of the analysts were able to provide an incident report at surprising precision in very limited time. Moreover, it seems that the results obtained from less and more skilled analysts are subtle. We are aware that it could be affected by the attack vector of the incident selected for the evaluation, but this unexpected finding is promising for further development.

Another interesting observation was made regarding the usage of proposed visual-analytics concepts and their combinations. We noticed different workflows in using the tool by different analysts. This finding indicates that the tool is sufficiently generic. It supports various approaches to the verification of hypotheses and collecting the evidence. Moreover, the results captured in Figure~\ref{fig:radar-chart} suggest that there could exist a favorite combination of analytical elements. For example, the analysts P2 and P5 used predominantly span windows with name filtering and a lot of push-pins, while P3 and P4 preferred span windows and clusters combined with only a few push-pins. Exploring such behavioral patterns would bring insight into analytical strategies. However, it requires a much deeper evaluation and analysis in future work.

Our work is still in progress. During the user study, we collected user feedback and requests for additional useful features. %In what follows, we summarize the most important enhancements inferred from the evaluation.

\emph{File system attributes management:} Multiple analysts forgot to cancel the per-attribute filtering during the investigation. This mistake led to false hypotheses and delay in the investigation. Emphasizing this filter or indicating that the \emph{List View} contains only entries with selected modifications are required.

\emph{Dealing with file system records:} The \emph{List View} is the primary source of information for investigators, and efficient manipulation with records has shown to be the key factor for the investigation process. In spite of searching, filtering, and smart navigation techniques implemented in the \emph{List View}, the analysts requested even more features for rapid navigation in the list. Especially, scrolling the list to a record by \texttt{CTRL+F} hotkey was missing. Currently, only highlighting and filtering out the data by the typed text is implemented in the tool. Also, the support of regular expressions and hiding records matching the typed text temporarily were required.
Complementary hierarchical views to the strictly temporal ordering of records, e.g., using treemaps to convey space requirements of file system parts, reveal anomalies, and navigate to them quickly, will be considered in the future work.

The current implementation of FIMETIS serves as an analytical and decision-making tool for file system metadata analysis (Figure~\ref{fig:workflow}). Although the evaluation proved the usefulness of the tool, users ask for the support of other parts of the investigation process as well. Reaching this goal requires making significant extensions to current functionality and then to the design. In what follows, we outline key requirements and their possible impact on visualizations and GUIs.

\emph{Incident report creation:} Incident reports are key outputs of the investigation process. As a lot of clues and pieces of the incident evidence appear during the interaction, it would be useful to use them for the report creation. Apart from online notes that have already been integrated into the new version of FIMETIS, investigators' feedback revealed possible changes in using bookmarks for this purpose. Currently, bookmarks are very simple. They are represented as push-pins referring to interesting records (points in time) and used for fast navigation (jumping to these records). Multiple analysts were asking for the possibility to distinguish between push-pins by color, tagging them, and making their own notes. Once the concept of bookmarks is moved from push-pins to advanced annotations, it would be possible to use them for the direct generation of incident reports or their parts.

\emph{Analysis of system logs:} File system metadata represents only one source of information for investigators. Other data sources, like system logs or network traffic data, are often available to provide a broader context. Especially so-called super-timelines, i.e., file system metadata merged with system logs, are often used for forensic investigation. Extending FIMETIS with system logs should be possible. Both types of data sources are time series. The proposed approaches to file system exploration seem to be reusable also for system logs. However, further research and evaluation are needed. It is especially necessary to balance between unified exploration, when an analyst uses both data types together, and distinguishing both contexts as they represent different knowledge with possibly different uncertainty.

\emph{Other information sources:} Ability to analyze other data sources like network traffic or memory snapshots are required by forensic investigators as well. However, they encode very different data with very different abstractions that require the application of specific visual-analysis techniques and concepts. Therefore, narrowly focused tools are designed that provide comprehensive visual-analytics interfaces~\cite{cappers2018}. Joining these information sources into a single "silver bullet" analytical tool can be counter-productive and going against the \textbf{R6} requirement.

We aim to address the aforementioned features and enhancements in future work. As the FIMETIS application is already used in practice for the investigation of real-world incidents (three incidents were successfully investigated by the security teams of Masaryk University and CESNET so far), we aim to utilize this experience to extend the functionality of the application further. Especially, we plan to introduce advanced user-defined clusters and the support of multiple timelines, e.g., records of system logs. These extensions will require changes in the current design and the development of new visual-analytic methods to cope with even bigger and more variable data.

\acknowledgments{
This work was supported by ERDF ``CyberSecurity, CyberCrime and Critical Information Infrastructures Center of Excellence'' (No. CZ.02.1.01/0.0/0.0/16\_019/0000822).}

\bibliographystyle{abbrv}

\bibliography{main}

\begin{thebibliography}{10}

\bibitem{Anderson2019}
R.~Anderson, C.~Barton, R.~Boehme, R.~Clayton, C.~Ganan, M.~Levi, T.~Moore, and
  M.~Vasek.
\newblock {{Measuring the Cost of Cybercrime}}.
\newblock In {\em Proceedings of the 18th Annual Workshop on the Economics of
  Information Security}, 2019.

\bibitem{Bangor2009}
A.~Bangor, P.~Kortum, and J.~Miller.
\newblock {Determining What Individual SUS Scores Mean: Adding an Adjective
  Rating Scale}.
\newblock {\em Journal of Usability Studies}, 4(3):114--123, May 2009.

\bibitem{boschetti2011}
A.~Boschetti, L.~Salgarelli, C.~Muelder, and K.-L. Ma.
\newblock {TVi: a visual querying system for network monitoring and anomaly
  detection}.
\newblock In {\em Proceedings of the 8th international symposium on
  visualization for cyber security}, pages 1--10, 2011.

\bibitem{buchholz2004}
F.~Buchholz and E.~Spafford.
\newblock {On the role of file system metadata in digital forensics}.
\newblock {\em Digital Investigation}, 1(4):298 -- 309, 2004.

\bibitem{buchholz2005}
F.~P. Buchholz and C.~Falk.
\newblock {Design and Implementation of Zeitline: a Forensic Timeline Editor}.
\newblock In {\em Proceedings of the fifth annual DRFWS Conference}, 2005.

\bibitem{cappers2018}
B.~Cappers.
\newblock {\em Interactive visualization of event logs for cybersecurity}.
\newblock PhD thesis, Department of Mathematics and Computer Science, Dec.
  2018.
\newblock Proefschrift.

\bibitem{carrier2005}
B.~Carrier.
\newblock {\em File System Forensic Analysis}.
\newblock Addison-Wesley Professional, 2005.

\bibitem{casey2009}
E.~Casey.
\newblock {\em Handbook of Digital Forensics and Investigation}.
\newblock Academic Press, Inc., 2009.

\bibitem{forensicsChallenges}
L.~Caviglione, S.~Wendzel, and W.~Mazurczyk.
\newblock {The Future of Digital Forensics: Challenges and the Road Ahead}.
\newblock {\em IEEE Security \& Privacy}, 15(6):12--17, 2017.

\bibitem{gartner}
{Gartner, Inc.}
\newblock {Gartner Forecasts Worldwide Information Security Spending to Exceed
  \$124 Billion in 2019}.
\newblock \url{https://muni.cz/go/c7a9e9}, August 2018.

\bibitem{gray2015}
C.~C. Gray, P.~D. Ritsos, and J.~C. Roberts.
\newblock {Contextual network navigation to provide situational awareness for
  network administrators}.
\newblock In {\em 2015 IEEE Symposium on Visualization for Cyber Security
  (VizSec)}, pages 1--8. IEEE, 2015.

\bibitem{hargreaves2012}
C.~Hargreaves and J.~Patterson.
\newblock An automated timeline reconstruction approach for digital forensic
  investigations.
\newblock {\em Digital Investigation}, 9:S69 -- S79, 2012.

\bibitem{heitzmann2008}
A.~Heitzmann, B.~Palazzi, C.~Papamanthou, and R.~Tamassia.
\newblock Effective visualization of file system access-control.
\newblock In {\em International Workshop on Visualization for Computer
  Security}, pages 18--25. Springer, 2008.

\bibitem{humphries2013}
C.~Humphries, N.~Prigent, C.~Bidan, and F.~Majorczyk.
\newblock {Elvis: Extensible log visualization}.
\newblock In {\em Proceedings of the Tenth Workshop on Visualization for Cyber
  Security}, pages 9--16, 2013.

\bibitem{humphries2014}
C.~Humphries, N.~Prigent, C.~Bidan, and F.~Majorczyk.
\newblock {Corgi: Combination, organization and reconstruction through
  graphical interactions}.
\newblock In {\em Proceedings of the Eleventh Workshop on Visualization for
  Cyber Security}, pages 57--64, 2014.

\bibitem{kalber2013}
S.~K{\"{a}}lber, A.~Dewald, and F.~C. Freiling.
\newblock {Forensic Application-Fingerprinting Based on File System Metadata}.
\newblock In {\em Proceedings of the IEEE 2013 Seventh International Conference
  on IT Security Incident Management and IT Forensics}, pages 98--112, 2013.

\bibitem{kavrestad2018fundamentals}
J.~K{\"a}vrestad.
\newblock {\em Fundamentals of Digital Forensics: Theory, Methods, and
  Real-Life Applications}.
\newblock Springer International Publishing, 2018.

\bibitem{leschke2013}
T.~R. Leschke and C.~Nicholas.
\newblock Change-link 2.0: a digital forensic tool for visualizing changes to
  shadow volume data.
\newblock In {\em Proceedings of the Tenth Workshop on Visualization for Cyber
  Security}, pages 17--24, 2013.

\bibitem{olsson2009}
J.~Olsson and M.~Boldt.
\newblock Computer forensic timeline visualization tool.
\newblock {\em Digital Investigation}, 6:S78 -- S87, 2009.
\newblock The Proceedings of the Ninth Annual DFRWS Conference.

\bibitem{roberts2007}
J.~C. Roberts.
\newblock State of the art: Coordinated \& multiple views in exploratory
  visualization.
\newblock In {\em Fifth International Conference on Coordinated and Multiple
  Views in Exploratory Visualization (CMV 2007)}, pages 61--71. IEEE, 2007.

\bibitem{rowe2011}
N.~Rowe and S.~Garfinkel.
\newblock {Finding Anomalous and Suspicious Files from Directory Metadata on a
  Large Corpus}.
\newblock In {\em Proceedings of the Digital Forensics and Cyber Crime}, 2011.

\bibitem{Sauro2011}
J.~Sauro.
\newblock {\em A Practical Guide to the System Usability Scale: Background,
  Benchmarks \& Best Practices}.
\newblock CreateSpace Independent Publishing Platform, 2011.

\bibitem{Sauro2009}
J.~Sauro and J.~S. Dumas.
\newblock Comparison of three one-question, post-task usability questionnaires.
\newblock In {\em Proceedings of the SIGCHI Conference on Human Factors in
  Computing Systems}, CHI '09, pages 1599--1608, New York, NY, USA, 2009. ACM.

\bibitem{scherr2008}
M.~Scherr.
\newblock Multiple and coordinated views in information visualization.
\newblock {\em Trends in Information Visualization}, 38:1--33, 2008.

\bibitem{Sedlmair2012}
M.~{Sedlmair}, M.~{Meyer}, and T.~{Munzner}.
\newblock Design study methodology: Reflections from the trenches and the
  stacks.
\newblock {\em IEEE Transactions on Visualization and Computer Graphics},
  18(12):2431--2440, Dec 2012.

\bibitem{stange2014}
J.-E. Stange, M.~D{\"o}rk, J.~Landstorfer, and R.~Wettach.
\newblock {Visual filter: graphical exploration of network security log files}.
\newblock In {\em Proceedings of the Eleventh Workshop on Visualization for
  Cyber Security}, pages 41--48, 2014.

\bibitem{ulmer2019}
A.~Ulmer, D.~Sessler, and J.~Kohlhammer.
\newblock Netcapvis: Web-based progressive visual analytics for network packet
  captures.
\newblock In {\em 2019 IEEE Symposium on Visualization for Cyber Security
  (VizSec)}, 2019.

\end{thebibliography}
\end{document}